%% file: mmWaveSDR.tex
\input{variables}
\newcommand\mydots{\hbox to 1em{.\hss.\hss.}}

\documentclass[conference,comsoc]{IEEEtran}
\IEEEoverridecommandlockouts

\usepackage{multicol}
\usepackage{lipsum}
\usepackage{mathtools}
\usepackage{amsthm}
\usepackage{acronym}
\usepackage[bookmarksopen=true]{hyperref}
\usepackage{stfloats}
\usepackage{amsfonts}
\usepackage{cite}
\usepackage{multirow}
\usepackage{bm}
\usepackage{amsmath,amssymb}
\usepackage{graphicx}
\usepackage[table,xcdraw]{xcolor}
\usepackage[caption=false,font=footnotesize]{subfig}
\usepackage[geometry]{ifsym}
\usepackage{array}
\usepackage[utf8]{inputenc}
\usepackage[T1]{fontenc}

\let\norm\undefined 
\DeclarePairedDelimiter\norm{\lVert}{\rVert}
\DeclarePairedDelimiter\inprod{\langle}{\rangle}

\usepackage{tikz}
\usepackage{lipsum}
\usetikzlibrary{calc,shadings,patterns}
\makeatletter
\tikzset{%
	remember picture with id/.style={%
		remember picture,
		overlay,
		save picture id=#1,
	},
	save picture id/.code={%
		\edef\pgf@temp{#1}%
		\immediate\write\pgfutil@auxout{%
			\noexpand\savepointas{\pgf@temp}{\pgfpictureid}}%
	},
	if picture id/.code args={#1#2#3}{%
		\@ifundefined{save@pt@#1}{%
			\pgfkeysalso{#3}%
		}{
			\pgfkeysalso{#2}%
		}
	}
}

\def\savepointas#1#2{%
	\expandafter\gdef\csname save@pt@#1\endcsname{#2}%
}

\def\tmk@labeldef#1,#2\@nil{%
	\def\tmk@label{#1}%
	\def\tmk@def{#2}%
}

\tikzdeclarecoordinatesystem{pic}{%
	\pgfutil@in@,{#1}%
	\ifpgfutil@in@%
	\tmk@labeldef#1\@nil
	\else
	\tmk@labeldef#1,(0pt,0pt)\@nil
	\fi
	\@ifundefined{save@pt@\tmk@label}{%
		\tikz@scan@one@point\pgfutil@firstofone\tmk@def
	}{%
		\pgfsys@getposition{\csname save@pt@\tmk@label\endcsname}\save@orig@pic%
		\pgfsys@getposition{\pgfpictureid}\save@this@pic%
		\pgf@process{\pgfpointorigin\save@this@pic}%
		\pgf@xa=\pgf@x
		\pgf@ya=\pgf@y
		\pgf@process{\pgfpointorigin\save@orig@pic}%
		\advance\pgf@x by -\pgf@xa
		\advance\pgf@y by -\pgf@ya
	}%
}

\makeatother

\newcounter{hatchNumber}
\DeclarePairedDelimiter\ceil{\lceil}{\rceil}
\DeclarePairedDelimiter\floor{\lfloor}{\rfloor}

\usepackage{cuted}
\makeatletter
\newif\ifAC@uppercase@first%
\def\Aclp#1{\AC@uppercase@firsttrue\aclp{#1}\AC@uppercase@firstfalse}%
\def\AC@aclp#1{%
	\ifcsname fn@#1@PL\endcsname%
	\ifAC@uppercase@first%
	\expandafter\expandafter\expandafter\MakeUppercase\csname fn@#1@PL\endcsname%
	\else%
	\csname fn@#1@PL\endcsname%
	\fi%
	\else%
	\AC@acl{#1}s%
	\fi%
}%
\def\Acp#1{\AC@uppercase@firsttrue\acp{#1}\AC@uppercase@firstfalse}%
\def\AC@acp#1{%
	\ifcsname fn@#1@PL\endcsname%
	\ifAC@uppercase@first%
	\expandafter\expandafter\expandafter\MakeUppercase\csname fn@#1@PL\endcsname%
	\else%
	\csname fn@#1@PL\endcsname%
	\fi%
	\else%
	\AC@ac{#1}s%
	\fi%
}%
\def\Acfp#1{\AC@uppercase@firsttrue\acfp{#1}\AC@uppercase@firstfalse}%
\def\AC@acfp#1{%
	\ifcsname fn@#1@PL\endcsname%
	\ifAC@uppercase@first%
	\expandafter\expandafter\expandafter\MakeUppercase\csname fn@#1@PL\endcsname%
	\else%
	\csname fn@#1@PL\endcsname%
	\fi%
	\else%
	\AC@acf{#1}s%
	\fi%
}%
\def\Acsp#1{\AC@uppercase@firsttrue\acsp{#1}\AC@uppercase@firstfalse}%
\def\AC@acsp#1{%
	\ifcsname fn@#1@PL\endcsname%
	\ifAC@uppercase@first%
	\expandafter\expandafter\expandafter\MakeUppercase\csname fn@#1@PL\endcsname%
	\else%
	\csname fn@#1@PL\endcsname%
	\fi%
	\else%
	\AC@acs{#1}s%
	\fi%
}%
\edef\AC@uppercase@write{\string\ifAC@uppercase@first\string\expandafter\string\MakeUppercase\string\fi\space}%
\def\AC@acrodef#1[#2]#3{%
	\@bsphack%
	\protected@write\@auxout{}{%
		\string\newacro{#1}[#2]{\AC@uppercase@write #3}%
	}\@esphack%
}%
\def\Acl#1{\AC@uppercase@firsttrue\acl{#1}\AC@uppercase@firstfalse}
\def\Acf#1{\AC@uppercase@firsttrue\acf{#1}\AC@uppercase@firstfalse}
\def\Ac#1{\AC@uppercase@firsttrue\ac{#1}\AC@uppercase@firstfalse}
\def\Acs#1{\AC@uppercase@firsttrue\acs{#1}\AC@uppercase@firstfalse}

\DeclareMathOperator{\sign}{sign}
\DeclareMathOperator{\rem}{{\tt rem}}
\DeclareMathOperator{\diag}{diag}
\def\diagOperator[#1]{\diag\left(#1\right)}
\def\signNormal[#1]{\sign\left(#1\right)}
\def\remOperator[#1][#2]{\rem(#1,#2)}
\def\signThreshold[#1][#2]{\sign_{#2}\left(#1\right)}

\input{acronyms}
\def\BibTeX{{\rm B\kern-.05em{\sc i\kern-.025em b}\kern-.08em
		T\kern-.1667em\lower.7ex\hbox{E}\kern-.125emX}}
\begin{document}

	\title{A Millimeter-Wave Software-Defined Radio for Wireless Experimentation
		\\
		\thanks{This research is supported in part by the NSF award CNS-1939334.}
		\author{
			\IEEEauthorblockN{
				Alphan \c{S}ahin\IEEEauthorrefmark{1}, 
				Mihail L. Sichitiu\IEEEauthorrefmark{2}, 
				\.{I}smail G\"{u}ven\c{c}\IEEEauthorrefmark{2}
			} \IEEEauthorblockA{\IEEEauthorrefmark{1}University of South Carolina, Columbia, and \IEEEauthorrefmark{2}North Carolina State University, Raleigh\\
				E-mails: 
				asahin@mailbox.sc.edu,  
				mlsichit@ncsu.edu, 
				iguvenc@ncsu.edu
			}
		} 
		
	}
	\maketitle

	\begin{abstract}
		In this study, we propose a low-cost and portable  \acl{mmWave} \ac{SDR} for wireless experimentation in the 60~GHz band. The proposed SDR uses Xilinx RFSoC2x2 and Sivers EVK06002 homodyne transceiver and provides a TCP/IP-based interface  for \ac{CC}-based baseband signal processing. To address the large difference between the processing speed of the \ac{CC} and the sample rate of \aclp{ADC}, we propose a method, called \ac{WTR}, where a hard-coded block detects a special trigger waveform to acquire a pre-determined number of IQ samples upon the detection. We also introduce a buffer mechanism to support discontinuous transmissions. By utilizing the WTR along with discontinuous transmissions, we conduct a beam sweeping experiment, where we evaluate 4096 beam pairs rapidly without compromising the flexibility of the CC-based processing. We also generate a dataset that allows one to calculate physical layer parameters such as \acl{SNR} and \acl{CFR} for a given pair of transmit and receive beam indices.
	\end{abstract}
	
	\section{Introduction}
	
	\acresetall
	
	\Ac{mmWave} communications is one of the key enablers for high-throughput systems by allowing directive links over a large bandwidth. While there has been extensive research activity for \ac{mmWave} systems, experimentation  in real-world environments is still a major challenge due to the lack of low-cost and portable mmWave \acp{SDR}, as compared to the ones for sub-6~GHz. In this work, we address this issue with a new \ac{SDR} solution that is low cost,  highly flexible and portable, and developed from commercial off-the-shelf components.
	
	In the literature, there is a substantial interest in developing mmWave SDRs. For example, in \cite{Zhao_2020}, an \ac{SDR} with multiple \acp{PAA} is proposed by hijacking a commercial IEEE 802.11ad radio. In \cite{Chen_2021}, mmWave transceivers from Sivers, IBM, and InterDigital are evaluated, which are paired with \acp{USRP} or Xilinx ZCU111 for COSMOS testbed.
	While Xilinx ZCU111 enables the testbed to support waveforms with a larger bandwidth, the USRP-based approach offers the flexibility of SDR.
	In \cite{Lacruz_2021}, four 60~GHz \acp{PAA} are connected to Xilinx ZCU111, where the design particularly focuses on the implementation of IEEE 802.11ad in the \ac{FPGA}. 
	In~\cite{Zavorka_2022}, ZCU111 is utilized with the discrete circuits for a wireless sensing application. 
	In \cite{Santhi_2022}, NI's mmWave solution along with SiBeam \acp{PAA} is considered for video transmission. In \cite{Sim_2022}, a full-duplex system is demonstrated by using custom boards. Although the proposed designs in~\cite{Zavorka_2022,Santhi_2022,Sim_2022} are complete solutions, the introduced platforms can be costly and not trivial to make them portable in practice. Also, they may not be accessible or constructed easily by many academic institutions.
	
	In this study, with the motivations of developing a portable, low-cost, and easy-to-construct mmWave SDR and building \acp{UAV} equipped with mmWave SDRs for NSF \ac{AERPAW} platform at North Carolina State University, we disclose a set of \ac{SDR} solutions.
	Our contributions in this work are as follows.
	
	{\bf Waveform-triggered reception:}   
	To  maintain the flexibility of the \ac{CC}-based baseband signal processing and address the large difference between the sample rate of the \ac{ADC}  and the \ac{CC}'s processing speed, we propose \ac{WTR}, where an \ac{IP} that detects a special trigger waveform and passes a pre-determined number of \ac{IQ} data samples followed by the trigger waveform to the \ac{PS}. Hence, \ac{WTR} paves the way for the reception of any waveform desired to be communicated between the SDRs while substantially reducing the load on the interface between \ac{CC} and \ac{SDR}. 
	
	{\bf A buffer method for discontinuous transmissions:} 
	By exploiting WTR, we introduce a buffer mechanism that automatically stores the \ac{IQ} data in a discontinuous manner. While this feature improves the resource utilization in the \ac{FPGA}, it enables fast \ac{CC}-based beam sweeping. With this feature, we conduct a beam sweeping experiment and create a dataset consisting of the IQ samples for a given TX-RX \ac{AWV} index pair \cite{dataIEEEDataPort}.

	\section{Millimeter-wave SDR Architecture}
	\begin{figure*}[t]
		\centering
		\subfloat[Block diagram.]{\includegraphics[width =5.0in]{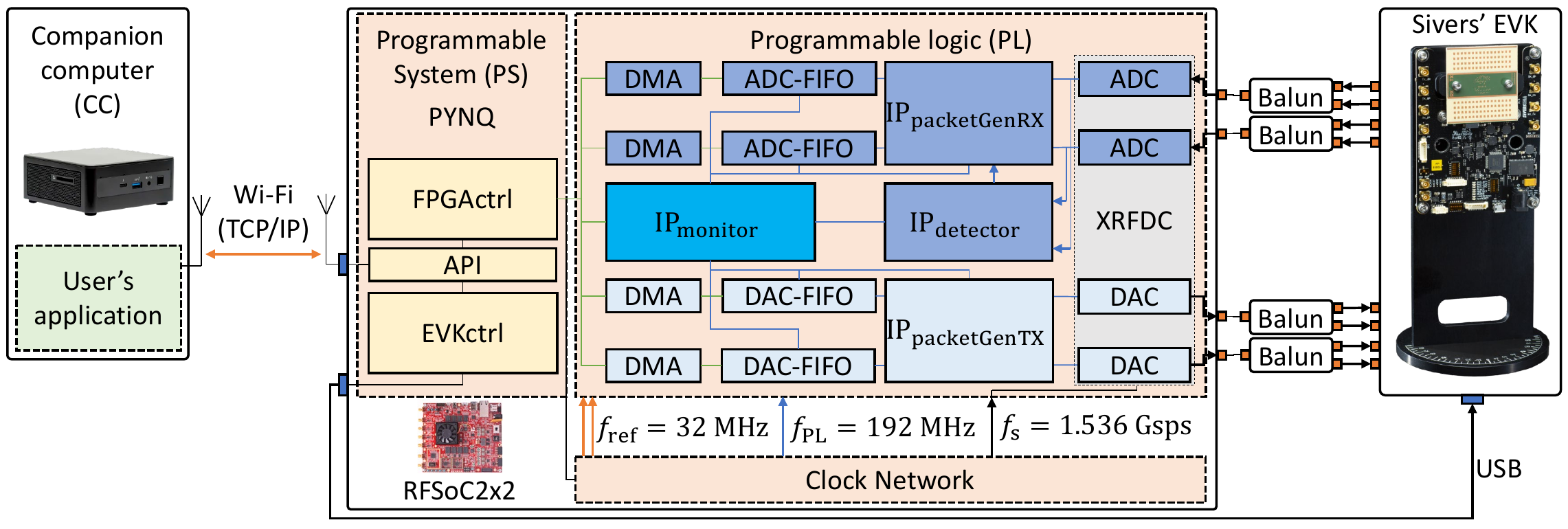}}~
		\subfloat[Implementation.]{\includegraphics[width =1.75in]{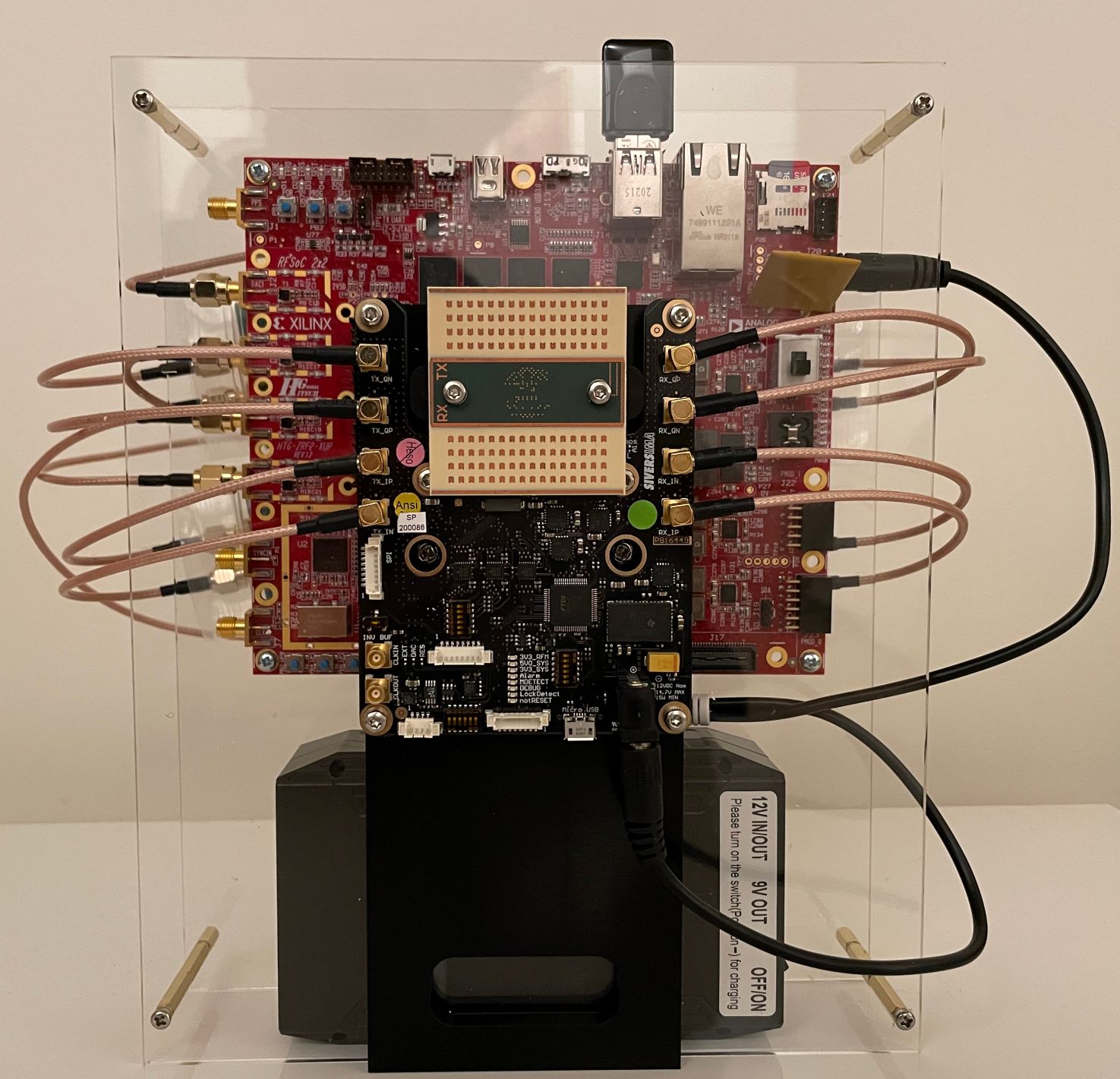}} 
		\caption{The architecture of the proposed SDR and the implemented SDR itself. The proposed radio can be used with a 12-V battery.}
		\label{fig:arhitecture}
		\vspace{-4mm}
	\end{figure*}

	
	For the proposed \ac{SDR}, we prioritize the flexibility so that an arbitrary waveform, e.g., the ones based on a standard such as IEEE 802.11ay/ad or 3GPP 5G \ac{NR} or a custom waveform for a particular experiment, can be used. To this end, our strategy is to implement a minimal feature set by taking the aspects of \ac{mmWave} communications into account. The proposed \ac{SDR} uses Sivers EVK06002 evaluation kit and RFSoC2x2 \ac{FPGA} board, and provides an \ac{API} for a \ac{CC}-based signal processing. All the constituents of the proposed \ac{SDR} are open source \cite{codeGitHub} and can be customized depending on the experiment requirements.
	
	EVK06002 is an evaluation kit that converts the continuous-time baseband in-phase and quadrature signals to the passband or vice versa. It hosts a BFM06010 RF module that integrates two \acp{PAA} for transmission and reception with a TRX BF/01 transceiver, i.e., a homodyne \ac{IQ} modulator/demodulator. Each \ac{PAA} provides 16 channels, where each channel is wired to $4$ patch antennas. The TRX BF/01 can be tuned within the range $57$-$71$~GHz and does not need an extra \ac{LO}. It can also store 64 custom \acp{AWV}, where the phases of in-phase and quadrature components for each channel can be controlled  between $-1$ to $1$ with the resolution of 6 bits. All the features of TRX BF/01 can be controlled via a \ac{USB} interface over an FTDI4232 chipset or the pin connections on EVK06002 with a custom \ac{SPI}.  The kit provides differential-ended in-phase and quadrature signals. We use four wide-band baluns, i.e., TI ADC-WB-BB, to convert them to single-end signals.
	
	
	%
	
	RFSoC2x2 is an \ac{FPGA} board that features Zynq UltraScale+ XCZU28DR. The \ac{FPGA} has built-in eight 12-bit \acp{ADC} and eight 14-bit \acp{DAC} with  the maximum sample rate of 4.096 Gsps and  6.554 Gsps, respectively.
	The board provides connections to two of the  \acp{ADC} and  two of the \acp{DAC} through SMA connectors. Hence, it allows one to synthesize signals up to 4.096 GHz bandwidth around the carrier with an \ac{IQ} modulator/demodulator. Also, the \ac{FPGA} integrates Arm Cortex-A53 64-bit quad-core processor
	and supports PYNQ: an open-source Linux-based system that facilitates the interaction between the \ac{FPGA} design, i.e., \ac{PL}, with a custom software, i.e., \ac{PS}. PYNQ runs on Cortex-A53 and supports Python language.

	In the proposed design, RFSoC2x2 is responsible for the following tasks:
	1) Generating the in-phase and quadrature signals based on the \ac{IQ} samples to be transmitted; 2) Acquiring a desired number of IQ samples by sampling the baseband in-phase and quadrature signals; 3) Configuring and controlling EVK06002 over a \ac{USB} port; and 4) Providing a TCP/IP-based \ac{API} for \ac{CC}. These tasks are managed by the objects $\FPGAcontroller$, $\EVKcontroller$, and  $\API$, running in PYNQ. For the first and second tasks, $\FPGAcontroller$  interacts with the \acp{IP} in the \ac{PL} via \ac{AXI}. It manages \ac{MTS} with a specific clock distribution, \ac{IQ} data acquisition or transmission, and \ac{WTR}. For the third task, $\EVKcontroller$ uses publicly available {\tt PyFtdi} library and provides a basic set of functions to read and set the registers of EVK06002. For the last task, $\API$ uses {\tt socket} library and establishes a TCP/IP connection with \ac{CC}. It provides a set of instructions to the \ac{CC} to control the radio. The proposed architecture and the implemented SDR are shown in \figurename~\ref{fig:arhitecture}.

	\section{Programmable Logic and System Design}
	\label{sec:pl}
\begin{figure}
		\centering
		\subfloat[$\txPacketGeneretor$.]{\includegraphics[width =3.2in]{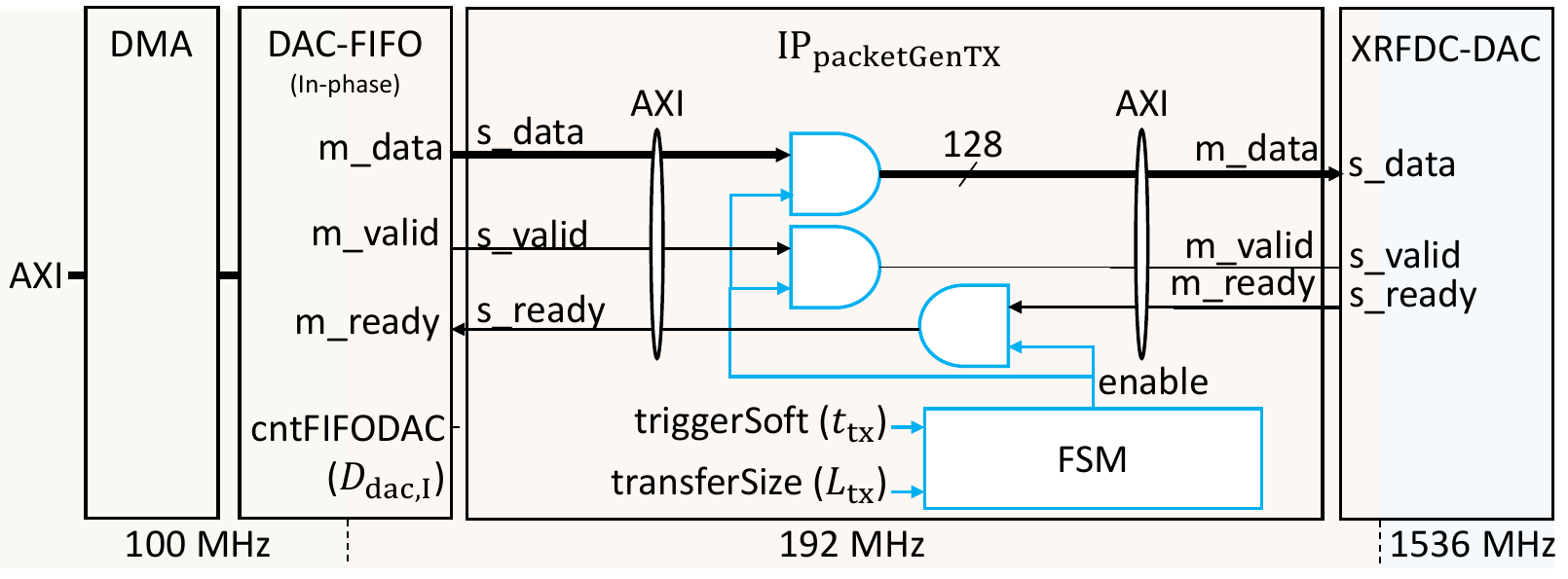}
			\label{subfig:txPacketGenerator}}\\
		\subfloat[$\rxPacketGeneretor$.]{\includegraphics[width =3.2in]{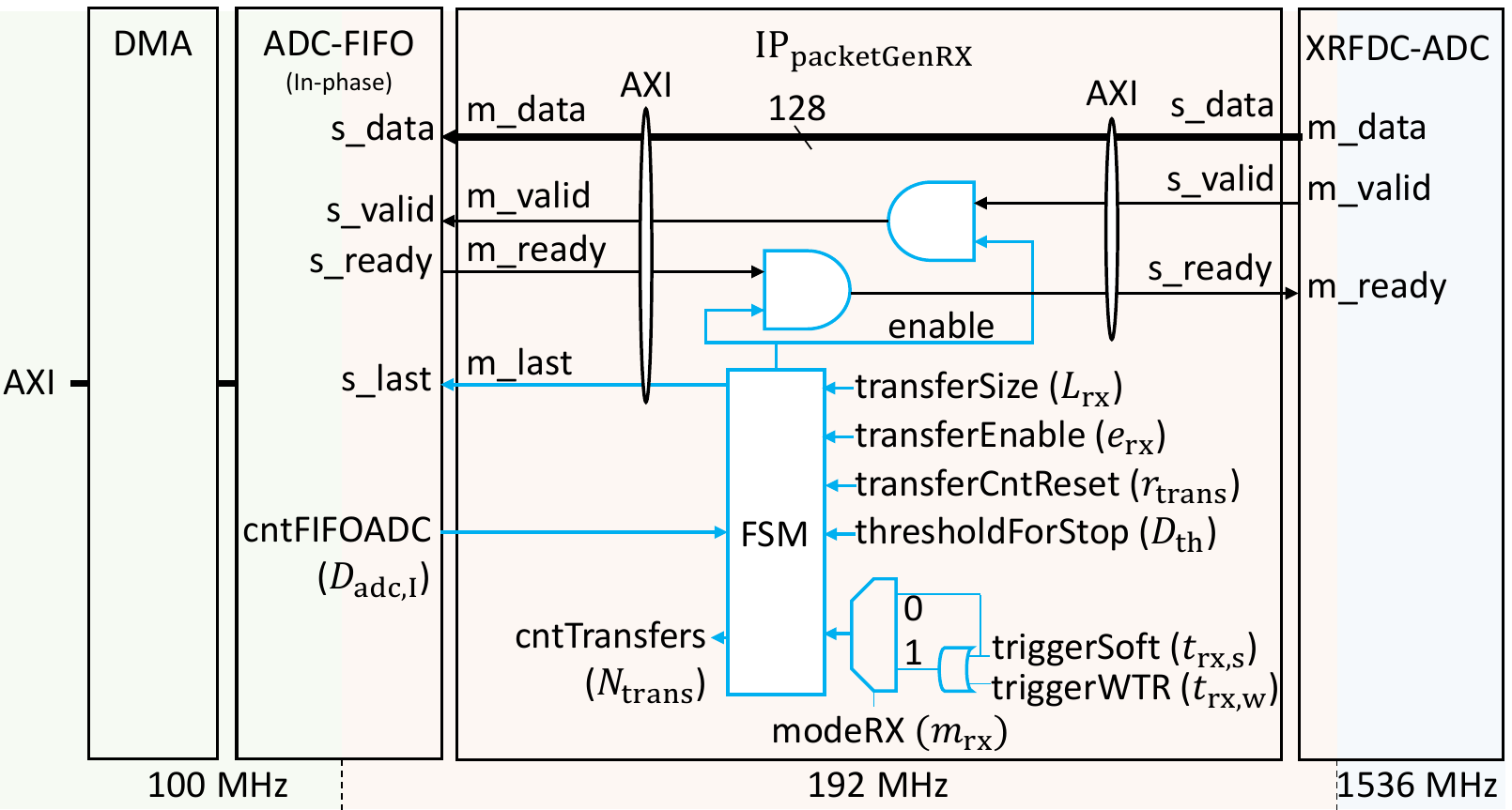}
			\label{subfig:rxPacketGenerator}}\\
		\subfloat[$\monitor$.]{\includegraphics[width =\figuresize-2.1in]{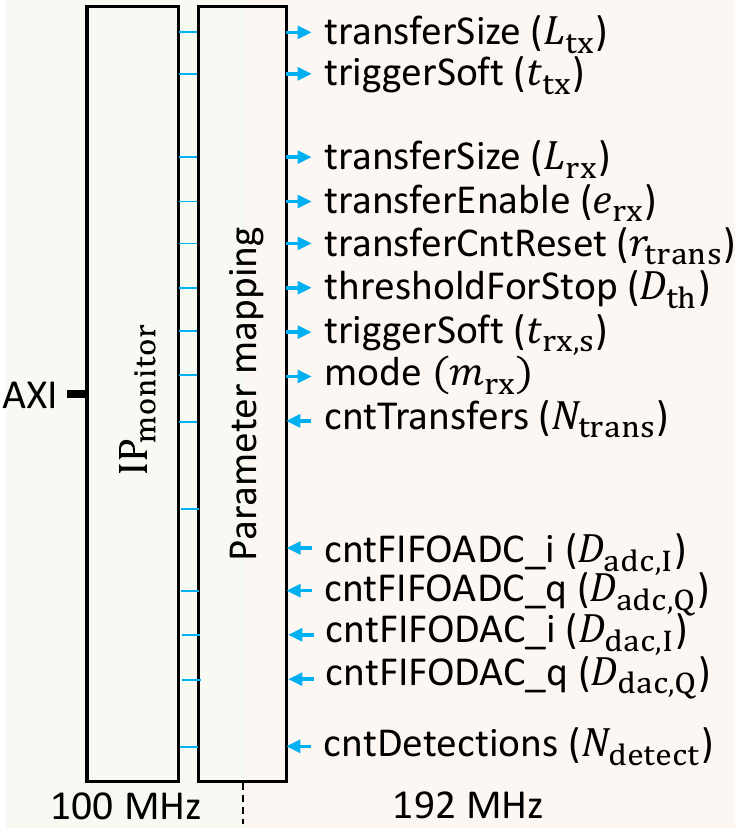}
			\label{subfig:monitor}}	
		\subfloat[$\detector$.]{\includegraphics[width =\figuresize-1.8in]{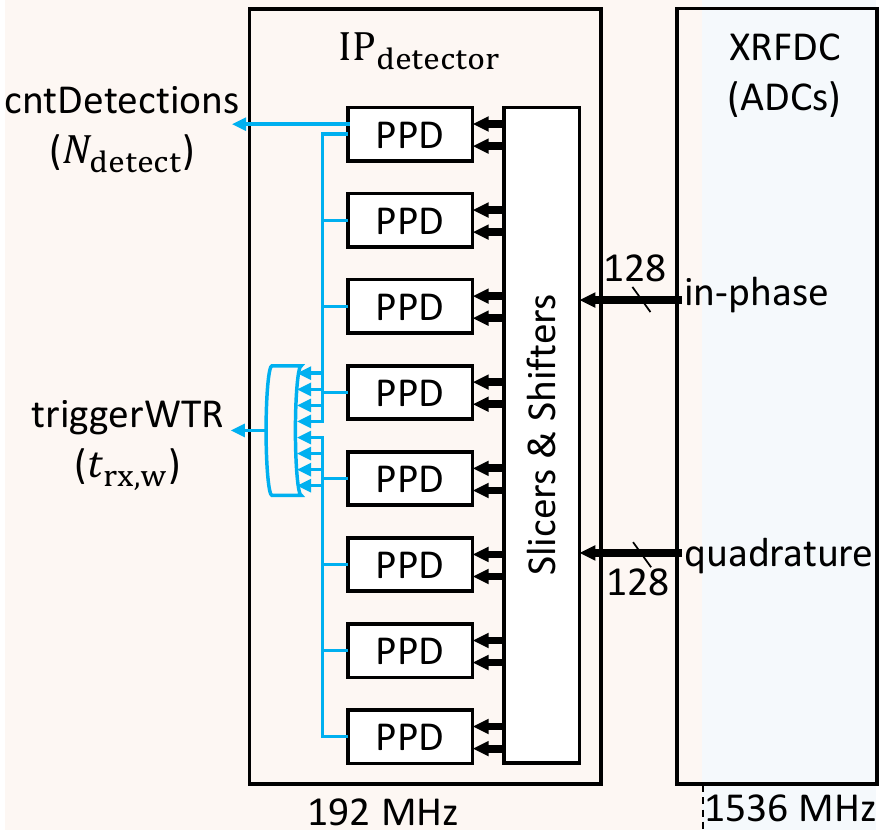}
			\label{subfig:detector}}~	
		\caption{The block diagrams of the developed IPs.}
		\label{fig:IPs}
		\vspace{-2mm}
	\end{figure}

	In this section, we introduce the developed \acp{IP} and discuss how they are utilized. 
	In addition to \ac{XRFDC} (contains both \acp{ADC} and \acp{DAC}), Xilinx \ac{AXI} \ac{FIFO}, and Xilinx \ac{AXI} \ac{DMA} blocks, we develop four main \acp{IP} to maximize the flexibility of the proposed SDR. $\txPacketGeneretor$ controls the \ac{IQ} data transfer from the DAC FIFOs to the \ac{XRFDC}. $\rxPacketGeneretor$ manages the \ac{IQ} data transfer from the \ac{XRFDC} to the ADC FIFOs. $\detector$ detects the trigger waveform for \ac{WTR}. $\monitor$ configures and reads the registers of  $\rxPacketGeneretor$, $\txPacketGeneretor$, $\detector$ and \acp{FIFO}. We use MATLAB HDL Coder Toolbox to develop  each \ac{IP}. The input/output ports of these IPs are shown in \figurename~\ref{fig:IPs} and their functions are described in the following subsections.

	We consider a clock distribution that allows the \acp{ADC} (or \acp{DAC}) on different tiles of \ac{XRFDC} to sample the in-phase and quadrature signals (or to convert the \ac{IQ} samples to continuous-time signals), {\em simultaneously}. To this end, we  use the \ac{MTS} feature of Xilinx RFSoCs and generate the two reference clocks (i.e., analog and digital reference clocks) at $\fref=32$~MHz, $\fpl=192$~MHz for \ac{PL}, and $\sampleRate=1.536$~GHz for the sample clocks.\footnote{Xilinx recommends the reference clocks for MTS to be less than 10~MHz. For our design, the MTS is maintained more accurately for 32~MHz.}
	The reference clocks are utilized to measure the latency and offset between the sampling instances of \acp{ADC} (or \acp{DAC}) based on Xilinx's guidelines on \ac{MTS}. We set the decimation and interpolation factors to 1. Hence, the sample rate of the \acp{ADC} and the \acp{DAC} are $1.536$~Gsps. As a result, the clock distribution allows the proposed \ac{SDR} to transmit or receive an arbitrary waveform with 1.536~GHz bandwidth maximum at Nyquist rate with an \ac{IQ} modulator.
	
	It is worth nothing that the \ac{XRFDC} operates with a clock rate that is $\conversionRate\triangleq\sampleRate/\fpl=8$ times faster than the one for the \ac{PL}. Hence, $\txPacketGeneretor$ and $\rxPacketGeneretor$ pushes or pulls $\conversionRate=8$ in-phase and  quadrature samples  concurrently for each \ac{PL} clock, where each sample is represented with 16 bits.
	If a higher $\sampleRate$ is needed for a specific experiment, either more parallel structures need to be introduced to the \ac{PL} or $\fpl$ needs to be increased to keep $\conversionRate$ constant, i.e., a trade-off between the \ac{FPGA} resources and the clock rate.

	\subsection{Transmission}
	For the transmission, we employ two \ac{AXI} \acp{FIFO}, i.e., DAC-FIFOs, for in-phase and quadrature samples, where their depths and widths are set to $\depthFIFO=2^{15}$ and $\widthFIFO=16\conversionRate$ bits, respectively. Hence, the proposed \ac{SDR} ensures the transmission of an \ac{IQ} data of length $2^{18}$ without an underflow. The steps for transmitting $\sizeIQdataTX$ \ac{IQ} samples  are as follows.
	
	\textbf{Step 1}: $\FPGAcontroller$ pads $\remOperator[\sizeIQdataTX][\conversionRate]$ zeroes to the \ac{IQ} data, where  $r=\remOperator[\sizeIQdataTX][\conversionRate]$ denotes the least positive remainder.
	
	\textbf{Step 2}: $\FPGAcontroller$ writes the padded \ac{IQ} data samples to the DAC-\acp{FIFO} via \acp{DMA}. 
	
	\textbf{Step 3}: $\FPGAcontroller$ sets the transfer size $\transferSizeTX$  as $\ceil{\sizeIQdataTX/\conversionRate}$.
	
	\textbf{Step 4}: $\FPGAcontroller$ triggers the transmission with the rising-edge of $\triggerSoftTX$, i.e., by changing its state from $0$ to $1$.
	
	\textbf{Step 5}: With the trigger, $\txPacketGeneretor$ enables \ac{XRFDC} to read $\conversionRate$ \ac{IQ} samples for $\transferSizeTX$ times from the DAC-FIFOs.
	
	In this study, all transmissions are initiated by the \ac{PS}. Also, $\FPGAcontroller$ sets or reads the registers $\transferSizeTX$ and $\triggerSoftTX$ via $\monitor$.

	\subsection{Reception}
	Similar to the transmission, two \ac{AXI} \acp{FIFO}, called ADC-FIFOs, for in-phase and quadrature samples are employed, where their depths and widths are $\depthFIFO=2^{15}$ and $\widthFIFO=16\conversionRate$ bits, respectively.  Hence, an IQ data of length $2^{18}$  can be acquired without an overflow. We introduce two modes, i.e., \ac{STR} and  \ac{WTR},  controlled by the flag $\modeRX$, as follows.
	
	\subsubsection{Software-triggered reception}
	If $\modeRX$ is set to $0$, the data acquisition is triggered by the \ac{PS}. This mode is useful for the measurement of existing signals in the environment. In this mode, $\sizeIQdataRX$ \ac{IQ} samples are received via the following steps:
	
	\textbf{Step 1}: $\FPGAcontroller$ sets the transfer size $\transferSizeRX$  as $\ceil{\sizeIQdataRX/\conversionRate}$ and enables the acquisition by setting the flag $\enableTransferRX$ to $1$.
	
	\textbf{Step 2}: $\FPGAcontroller$ triggers the acquisition with the rising-edge of the flag $\triggerSoftRX$.
	
	\textbf{Step 3}: With the trigger, $\rxPacketGeneretor$ enables ADC-\acp{FIFO} to pull $\conversionRate$  \ac{IQ} samples for $\transferSizeRX$ times.
	
	\textbf{Step 4}: $\rxPacketGeneretor$ marks the last sample via mlast flag.
	
	\textbf{Step 5}: $\FPGAcontroller$  reads the \ac{IQ}  samples from the ADC-\acp{FIFO} via \acp{DMA}. 
	
	\textbf{Step 6}: $\FPGAcontroller$ drops the last $\remOperator[\sizeIQdataTX][\conversionRate]$ samples to match with $\sizeIQdataRX$.

	\subsubsection{Waveform-triggered reception}
	\label{subsec:syncWaveform}
	For $\modeRX=1$, the acquisition is intended\footnote{For $\modeRX=1$, the acquisition can be still triggered by the PS via $\triggerSoftRX$ since this feature can be useful for certain applications or tests.} to be triggered by $\detector$ upon the detection of waveform $\ssSync$. The steps for WTR are as follows: 
	
	\textbf{Step 1}: $\FPGAcontroller$ configures the transfer size $\transferSizeRX$ for each transfer and set the threshold $\thresholdForStopRX$ to avoid overflow. 
	
	\textbf{Step 2}: The $\detector$ constantly searches for the trigger waveform with a set of \acp{PPD}. If one of the \acp{PPD} detects the trigger waveform $\ssSync$, it rises $\triggerWTR$. 
	
	\textbf{Step 3}: With the trigger, if there is enough room in the ADC-\acp{FIFO}, $\rxPacketGeneretor$ enables ADC-\acp{FIFO} to pull $\conversionRate$ \ac{IQ} samples for $\transferSizeRX$ times. Hence, the $\transferSizeRX\conversionRate$ \ac{IQ} samples following upon the detection instant are pulled to the ADC-FIFOs. 
	
	\textbf{Step 4}: $\rxPacketGeneretor$ marks the last sample and increases $\cntTransferRX$ by $1$ to indicate the number of transfers to the \ac{PS}.
	
	\textbf{Step 5}: $\FPGAcontroller$ reads $\cntTransferRX$. If $\cntTransferRX>0$, $\FPGAcontroller$ can read the $\cntTransferRX\times\transferSizeRX\conversionRate$ \ac{IQ} data samples from the ADC-\acp{FIFO} via \acp{DMA}. 
	
	
	As compared to \ac{STR}, the main difference of \ac{WTR} is that the trigger source is the \ac{PL}. Hence, the data flow needs to be managed by the \ac{PL}. In this study, the \ac{PS} sets $\thresholdForStopRX=\transferSizeRX\times\floor{\depthFIFO/\transferSizeRX}$ and the \ac{PL} checks if $\cntFIFOADC<\thresholdForStopRX$ holds to ensure that there is enough room in the ADC-FIFOs for the next transfer, where $\cntFIFOADC$ is the read counter of one of the ADC-\acp{FIFO}. Note that $\FPGAcontroller$ can reset $\cntTransferRX$ via $\transferCntResetRX$ and flush the ADC-FIFOs at any time. The registers $\transferSizeRX$,  $\triggerSoftRX$, $\modeRX$, $\enableTransferRX$, $\transferCntResetRX$,  $\thresholdForStopRX$, and $\cntTransferRX$  are set or read via $\monitor$.
	
	\paragraph{Buffer mechanism for discontinuous transmissions}
	One of the unique features of the \ac{WTR} is that it allows discontinuous transmissions, i.e., it enables ADC-\acp{FIFO} to store $\cntTransferRX=\floor{\depthFIFO/\transferSizeRX}$ transfers in the ADC-FIFOs, where the receptions depend on the transmission instants. This feature is particularly useful for increasing the speed for \ac{CC}-based beam sweeping. For example, consider a scenario where the transmitter transmits a set of IQ samples of length $1024$ to identify the best \ac{AWV} in a beambook of size $64$. Let the IQ samples encode the utilized \ac{AWV} index. By  transmitting the corresponding waveform along with the trigger waveform at different times and using \ac{WTR} at the receiver for $\transferSizeRX\ge256$, among 64 transmissions, only the ones detected by the \acp{PPD} are written to the ADC-FIFOs. Once the transmissions are completed, the \ac{CC} at the receiver side reads $\cntTransferRX$ (i.e., Step 5 in Section~\ref{subsec:syncWaveform}) to identify how many successful receptions (or transfers) is occurred. By pulling and processing the corresponding IQ data samples, the \ac{CC} can decode the beam indices and identify the best \ac{AWV} at the transmitter without continuously monitoring the IQ samples. We use this scenario for our experiment as discussed in Section~\ref{sec:numericalResults}.

	\paragraph{Trigger waveform and detector}
	We design the trigger waveform  $\ssSync$ and its detection based on the strategy in \cite{sahinGCdemo_2022}. In this method, the sequence  $\ssSync$ is a \ac{SC} waveform with the roll-off factor of $\rollOff=0.5$ synthesized by upsampling a repeated \ac{BPSK} modulated sequence, i.e., $2[\golaySequnce,~\golaySequnce,~\golaySequnce,~\golaySequnce]-1$, by a factor of $\Nup=4$ and passing it through a \ac{RRC} filter,
	where $\golaySequnce=[\golaySequnceEle[0],\mydots,\golaySequnceEle[31]]\in\realNumbers^{1\times32}$ is a binary Golay sequence. As a result, the null-to-null bandwidth of  $\ssSync$ is equal to $(1+\rollOff)/\Nup\times\sampleRate=3/8\times\sampleRate=
	576$~MHz. In \cite{sahinGCdemo_2022}, the design of $\ssSync$ is motivated as follows: 
	1) A cross-correlation operation with this waveform can be realized by using an approximate waveform where its samples are either $1$ or $-1$. This feature leads to better utilization of \ac{FPGA} resources since the multiplications can be reduced to  additions or subtractions. 2) By detecting the presence of shorter sequence $\golaySequnce$  back-to-back four times, the distortion due to the \acl{CFO} can be circumvented. 3) A smaller dynamic range with \ac{SC} waveform requires less power back-off. The metric that is used for the detection of $\golaySequnce$ can be expressed as:
	\begin{align}
		\metric[\indexSample]\triangleq\frac{1}{\norm{\signalTwo}^2}\frac{|\xcorr[\indexSample]|^2}{|\acorr[\indexSample]|^2}=\frac{1}{\norm{\signalTwo}^2}\frac{\inprod{\signalOne[\indexSample],\signalTwo}^2}{\inprod{\signalOne[\indexSample],\signalOne[\indexSample]}^2} = \frac{\inprod{\signalOne[\indexSample],\signalTwo}^2/2^{14}}{\norm{\signalOne[\indexSample]}^2}~,
		\label{eq:metric}
	\end{align}
	where $\signalTwo=[\signalTwoEle[0],\signalTwoEle[1],\mydots,\signalTwoEle[127]]$ is based on the aforementioned approximate \ac{SC} waveform with the rectangular filter and equal to $\signalTwo=2[\golaySequnceEle[31],\golaySequnceEle[31],\golaySequnceEle[31],\golaySequnceEle[31],\golaySequnceEle[30],\golaySequnceEle[30],\golaySequnceEle[30],\golaySequnceEle[30],\mydots,\golaySequnceEle[0],\golaySequnceEle[0],\golaySequnceEle[0],\golaySequnceEle[0]]-1$ for $\Nup=4$ and $\signalOne[\indexSample]=[\iqRXinput[\indexSample],\iqRXinput[\indexSample-1],\mydots,\iqRXinput[\indexSample-127]]$, where $\iqRXinput[\indexSample]$ is the $\indexSample$th received \ac{IQ} sample. A \ac{PPD} declares a detection  if $\metric[\indexSample]$ is larger than 1/4 for four times with 128 samples apart. We also implement a counter for  the first PPD to  monitor the number of detection events, i.e., $\cntDetect$, for test purposes.  
	
\begin{figure}[t]
		\centering
		{\includegraphics[width =3.5in]{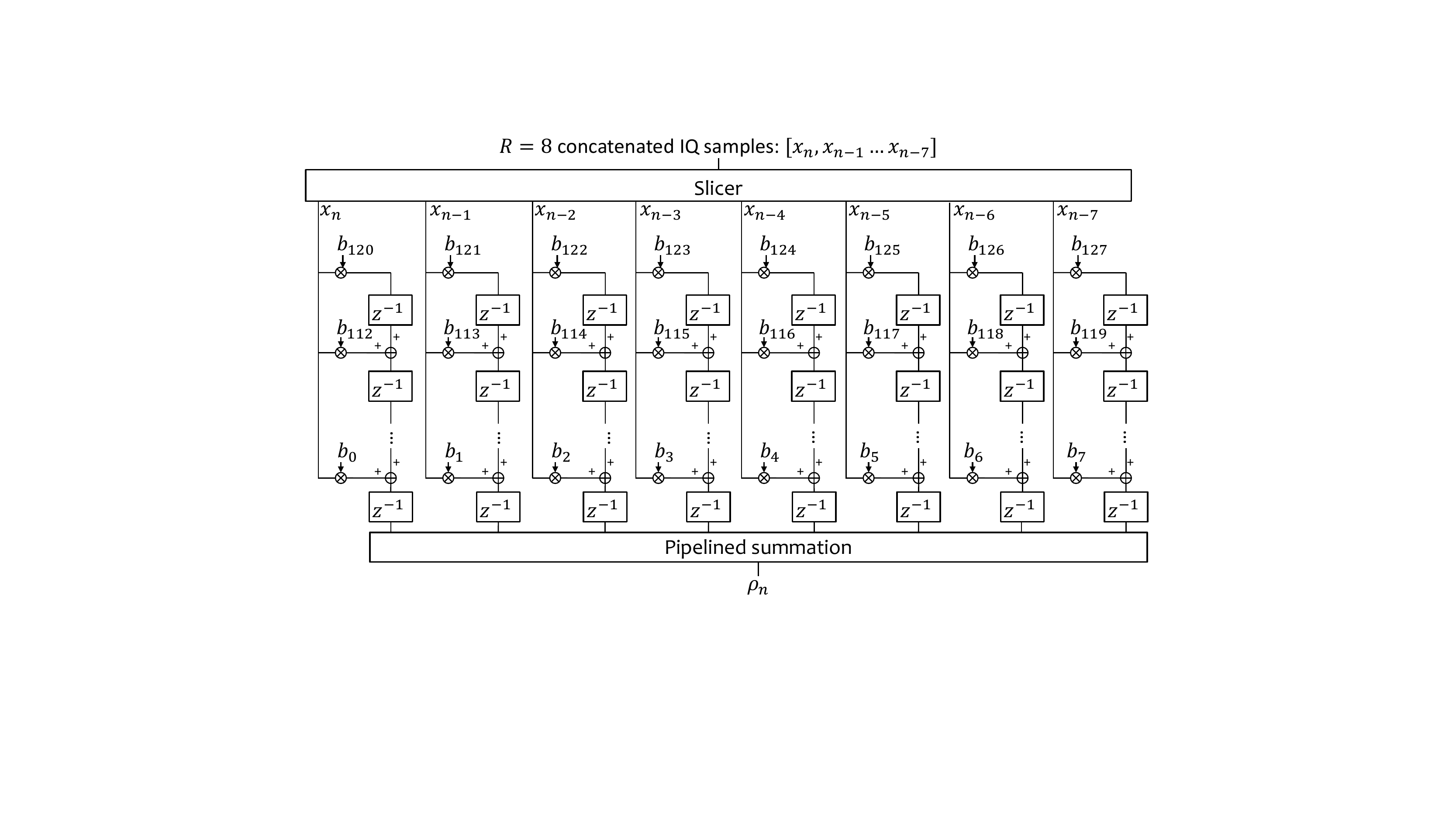}
		} 
		\vspace{-6mm}
		\caption{Cross-correlation implementation in a PPD.}
		\label{fig:ppd}
	\end{figure}

	XRFDC provides $\conversionRate$  in-phase and  quadrature samples concurrently for each \ac{PL} clock due to the difference between the \ac{PL} clock and ADC sample rate. Hence, we implement the cross-correlation operation in \eqref{eq:metric}, i.e., $\inprod{\signalOne[\indexSample],\signalTwo}$, as a \ac{FIR} filter and exploit the following identity to calculate the result for every other $\conversionRate=8$ samples:
	\begin{align}
		\xcorr[\indexSample] = \sum_{\indexCorr=0}^{127}\signalTwoEle[\indexCorr]\iqRXinput[\indexSample-\indexCorr]
		=\sum_{\indexParallel=0}^{7}
		\xcorrParallel[\indexSample][\indexParallel]~,
		\label{eq:firAll}
	\end{align}
	for $\xcorrParallel[\indexSample][\indexParallel]=\sum_{\indexCorr=0}^{15}\signalTwoEle[8\indexCorr+\indexParallel]\iqRXinput[\indexSample-8\indexCorr-\indexParallel]$. As can be seen from \eqref{eq:firAll}, for each PL clock, $\xcorr[\indexSample]$ can be obtained by summing the outputs of $8$ sub-\ac{FIR} filters, i.e., $\xcorrParallel[\indexSample][\indexParallel]$, where the input of the $\indexParallel$th sub-filter is the $\indexParallel$th sample of $\conversionRate$~IQ samples provided by XRFDC. We implement the \ac{FIR} filter by using the transposed form of each sub-filter with pipelining, as illustrated in \figurename~\ref{fig:ppd}. To detect $\ssSync$, we use $\conversionRate=8$ parallel \acp{PPD}  that operate on different lags. If one of the PPDs detect  $\ssSync$, the flag $\triggerWTR$ is raised. 
	
	\def\omitTheAPIcommands{1}
	\subsection{Application-programming interface}
	\label{subsec:api}
	The \ac{CC} interacts with the proposed SDR  via the instruction set defined in the $\API$ object. We define two TCP/IP ports for control and data. While control port is utilized to transfer commands and acknowledgments, the data port is utilized to exchange \ac{IQ} samples. With the developed \ac{API}, the \acp{AWV}, \ac{AWV} indices, carrier frequency, and gains can be controlled by the CC. 
	\if \omitTheAPIcommands 1
	The details related to the instruction set in the API are omitted due to the page limitation.
	\else
	The defined commands are given in \tablename~\ref{tab:commands}. 
	\begin{table*}[]
		\centering
		\label{tab:commands}
		\caption{Defined commands in the API.}
		\begin{tabular}{l|l|c|c}
			Command                       & \multicolumn{1}{c|}{Description}                                                                                             & \begin{tabular}[c]{@{}c@{}}Response \\ @ 8080\end{tabular} & \begin{tabular}[c]{@{}c@{}}Response\\ @8081\end{tabular}                                 \\ \hline
			setupReception                & \begin{tabular}[c]{@{}l@{}}Sets up the mode of the reception \\ Example: setupReception $\modeRX,\sizeIQdataRX$\end{tabular} & Read success/error                                         & N/A                                                                                      \\ \hline
			receiveIQSamples              & \begin{tabular}[c]{@{}l@{}}Receives IQ samples\\ Example: receiveIQSamples $\cntTransferRX$, timeOut\end{tabular}            & Read success/error                                         & \begin{tabular}[c]{@{}c@{}}Read $\cntTransferRX\sizeIQdataRX$\\  IQ samples\end{tabular} \\ \hline
			transmitIQSamples             & \begin{tabular}[c]{@{}l@{}}Transmit IQ samples\\ Example: transmitIQSamples $\sizeIQdataTX$\end{tabular}                     & Read success/error                                         & \begin{tabular}[c]{@{}c@{}}Write $\sizeIQdataTX$ \\ IQ samples\end{tabular}              \\ \hline
			getNumberOfAvailableTransfers & \begin{tabular}[c]{@{}l@{}}Returns $\cntTransferRX$.\\ Example: getNumberOfAvailableTransfers\end{tabular}                   & Read $\cntTransferRX$/error                                & N/A                                                                                      \\ \hline
			setTransferEnableRXFlag       & \begin{tabular}[c]{@{}l@{}}Set $\enableTransferRX$\\ Example: setTransferEnableRXFlag 1\end{tabular}                         & Read success/error                                         & N/A                                                                                      \\ \hline
			getFGPABitStreamFileName      & \begin{tabular}[c]{@{}l@{}}Returns the loaded bitstream\\ Example: getFGPABitStreamFileName\end{tabular}                     & Read filename/error                                        & N/A                                                                                      \\ \hline
			getRegisters                  & \begin{tabular}[c]{@{}l@{}}Returns all the register in $\monitor$\\ Example: getRegisters\end{tabular}                       & Read all registers                                         & N/A                                                                                      \\ \hline
			getBeamIndexTX                & \begin{tabular}[c]{@{}l@{}}Returns the current TX beam index\\ Example: getBeamIndexTX\end{tabular}                          & Read beam index/error                                      & N/A                                                                                      \\ \hline
			setBeamIndexTX                & \begin{tabular}[c]{@{}l@{}}Set the TX beam index\\ Example: setBeamIndexTX 32\end{tabular}                                   & Read success/error                                         & N/A                                                                                      \\ \hline
			getBeamIndexRX                & \begin{tabular}[c]{@{}l@{}}Returns the current RX beam index\\ Example: getBeamIndexRX\end{tabular}                          & Read beam index/error                                      & N/A                                                                                      \\ \hline
			setBeamIndexRX                & \begin{tabular}[c]{@{}l@{}}Set the RX beam index\\ Example: setBeamIndexRX 63\end{tabular}                                   & Read success/error                                         & N/A                                                                                      \\ \hline
			getModeSiver                  & \begin{tabular}[c]{@{}l@{}}Returns mode of EVKs\\ Example: getModeSiver\end{tabular}                                         & Read mode/error                                            & N/A                                                                                      \\ \hline
			setModeSiver                  & \begin{tabular}[c]{@{}l@{}}Set the mode of EVKs (e.g., RXen0\_TXen0)\\ Example: getModeSiver RXen1\_TXen0\end{tabular}       & Read mode/error                                            & N/A                                                                                      \\ \hline
			getGainRX                     & \begin{tabular}[c]{@{}l@{}}Returns the RX gain registers\\ Example: getGainRX\end{tabular}                                   & Read registers/error                                       & N/A                                                                                      \\ \hline
			setGainRX                     & \begin{tabular}[c]{@{}l@{}}Set the RX gain registers\\ Example: setGainRX FF 77 FF\end{tabular}                              & Read success/error                                         & N/A                                                                                      \\ \hline
			getGainTX                     & \begin{tabular}[c]{@{}l@{}}Returns the TX gain registers\\ Example: getGainTX\end{tabular}                                   & Read registers/error                                       & N/A                                                                                      \\ \hline
			setGainTX                     & \begin{tabular}[c]{@{}l@{}}Returns the TX gain registers\\ Example: setGainTX 44 FF FF\end{tabular}                          & Read success/error                                         & N/A                                                                                      \\ \hline
			getCarrierFrequency           & \begin{tabular}[c]{@{}l@{}}Get the current carrier frequency\\ Example: getCarrierFrequency\end{tabular}                     & Read frequency/error                                       & N/A                                                                                      \\ \hline
			setCarrierFrequency           & \begin{tabular}[c]{@{}l@{}}Set the carrier frequency\\ Example: setCarrierFrequency 60.4e9\end{tabular}                      & Read success/error                                         & N/A                                                                                      \\ \hline
		\end{tabular}
	\end{table*}
	\fi

	\section{Beam Sweeping Experiment}
	
	\label{sec:numericalResults}

\begin{figure}[t]
		\centering
		\subfloat[Experiment environment.]{\includegraphics[width =2.9in]{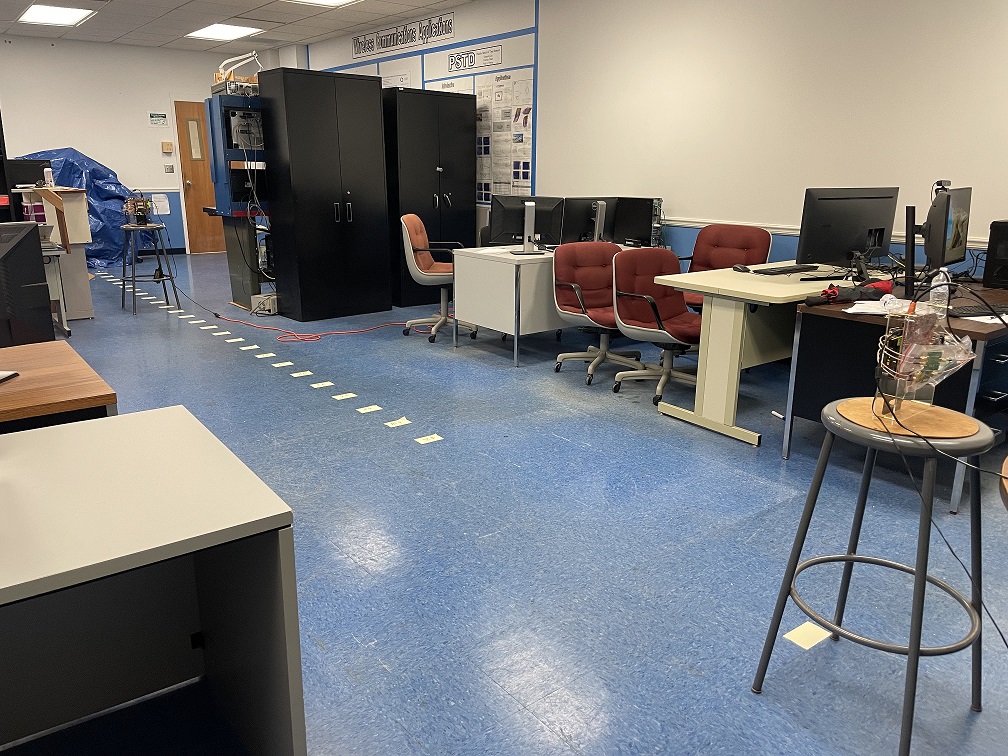}\label{subfig:env}}\\	
		\subfloat[Connections in the experiment.]{\includegraphics[width =2.8in]{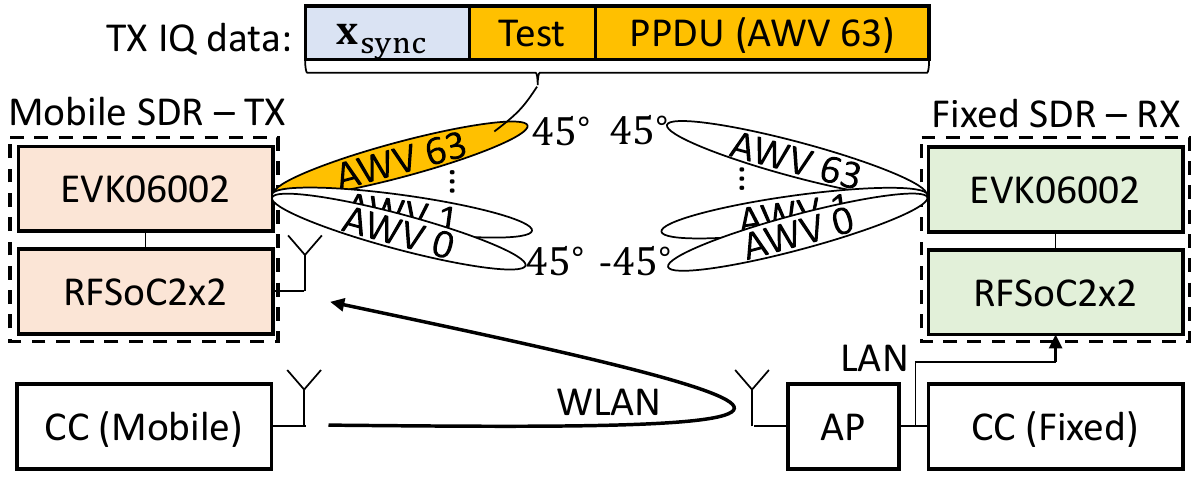}\label{subfig:connections}}
		\caption{Experiment setup.}
		\label{fig:experiment}
		\vspace{-4mm}
	\end{figure}

	In this study, we demonstrate the \ac{WTR} and the buffering method for discontinuous transmission with a beam sweeping experiment in an indoor office environment.  
	In the experiment,  we use one fixed SDR and a mobile SDR, where the SDRs face each other, as can be seen in  
	\figurename~\ref{fig:experiment}\subref{subfig:env}. Each \ac{SDR} is controlled by a CC over an \ac{AP}. As illustrated in \figurename~\ref{fig:experiment}\subref{subfig:connections}, the mobile SDR is controlled over the \ac{WLAN} since a wireless control allows us to adjust the mobile SDR's position conveniently in the experiment. We reduce the link distance between the SDRs, denoted by $\linkDistance$, from  9.75~m to 2.44~m with a spacing of 12" (i.e., 25 locations). We use the default customizable set of 64 \acp{AWV}  for $\carrierFrequency=60.48$~GHz, provided by Sivers, sweeping the azimuth between $-45^\circ$ and $45^\circ$ uniformly (i.e., side wall to side wall).
	Hence, in total, there exist 4096 TX-RX AWV index pairs one can choose for the mmWave link. Our goal with the experiment is to evaluate the \ac{SNR} for each pair at different locations for  $\carrierFrequency\in\{60.48,65.34\}$~GHz.
	\begin{figure}[t]
		\centering
		\subfloat[The IQ samples. The received test waveform and PPDU are shown. The detected TX AWV index is 31 for both signals.]{{\includegraphics[width =3.2in]{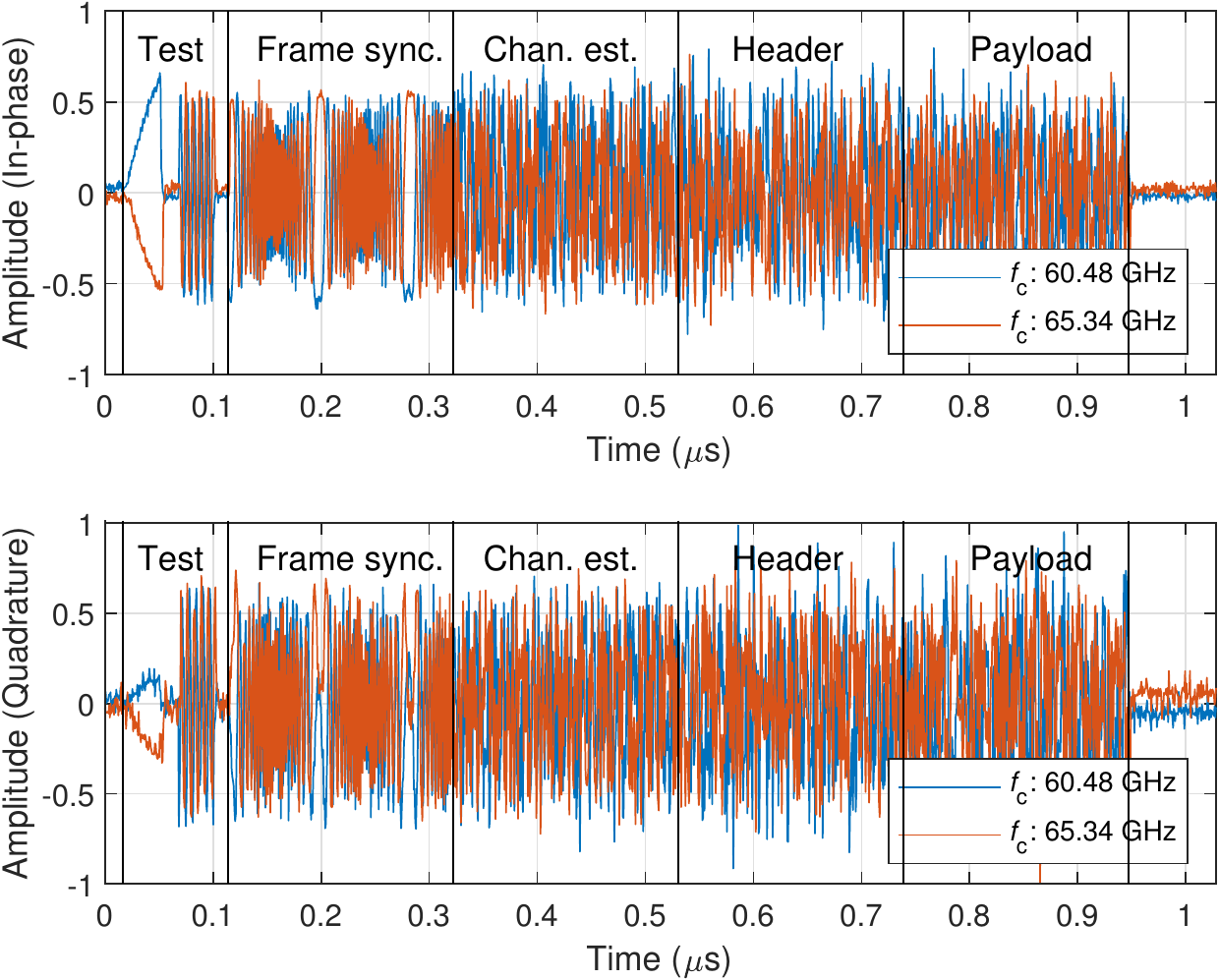}} }\\
		\subfloat[The measured CFR and CIR. 
		]{{\includegraphics[width =3.2in]{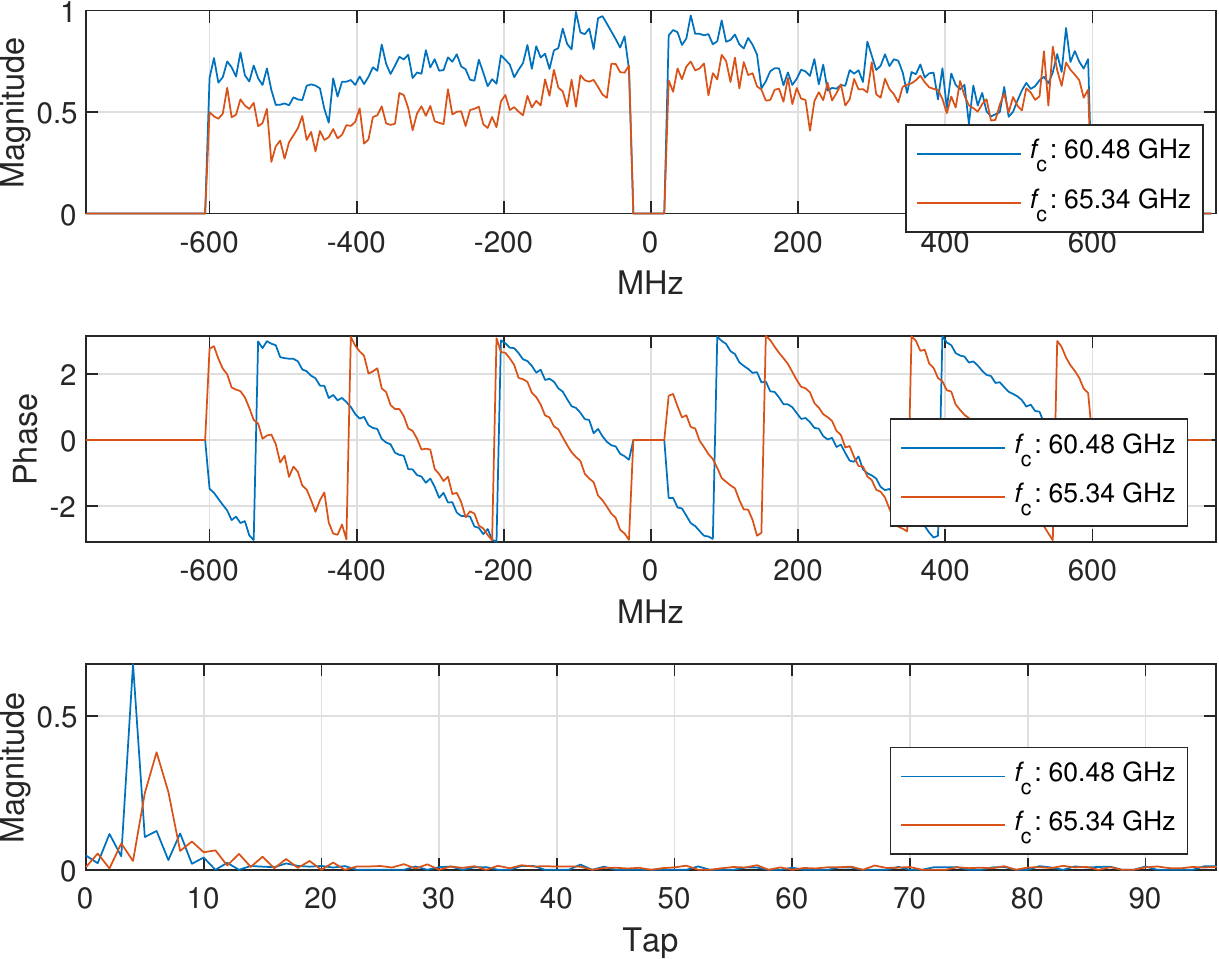}} }
		\caption{The received IQ samples for two transfers and their analyses. }
		\label{fig:dataPoint}
		\vspace{-2mm}
	\end{figure}
	
	We implement the following routine to calculate the SNR for a given TX-RX AWV index pair. The CC of the mobile SDR first sets the TX AWV index. It then generates an \acl{OFDM}-based \ac{PPDU} \cite{sahinGCdemo_2022}, where the data bits indicate the utilized TX AWV index. The PPDU consists of four \ac{OFDM} symbols for synchronization, channel estimation, header data, and payload and its length is $1280$ complex samples. The CC transmits the PPDU along with a test waveform and the trigger waveform $\ssSync$, where the test waveform consists of a ramp waveform ($50$ samples), zero samples ($25$ samples), a tone ($50$ samples), and zero samples ($25$ samples) for evaluating potential impairments, {\em visually}. After the transmission, the CC increases the TX AWV index and repeats the aforementioned announcement procedure. In our experiment, the CC completes the announcements of 64 TX AWV indices in less than 2~sec. The fixed SDR utilizes \ac{WTR}. The corresponding CC first sets the RX AWV index. It waits for 2~sec and reads $\cntTransferRX$. 
	For each transfer, it then pulls the corresponding IQ samples (1580 samples) and tries to decode the PPDU. If the decoding is successful, the CC detects announced TX AWV index and measures the SNR. It is worth noting we do not implement any exhaustive correlation in the CC to find the transmitted PPDU, thanks to the WTR. 
	With this procedure, we record the IQ data for all transfers for a given RX AWV index, location, and $\carrierFrequency$ and generate a dataset.
	

	\begin{figure*}[t]
		\centering
		\subfloat[$\carrierFrequency= 60.48$~GHz.]{{\includegraphics[width =3.2in]{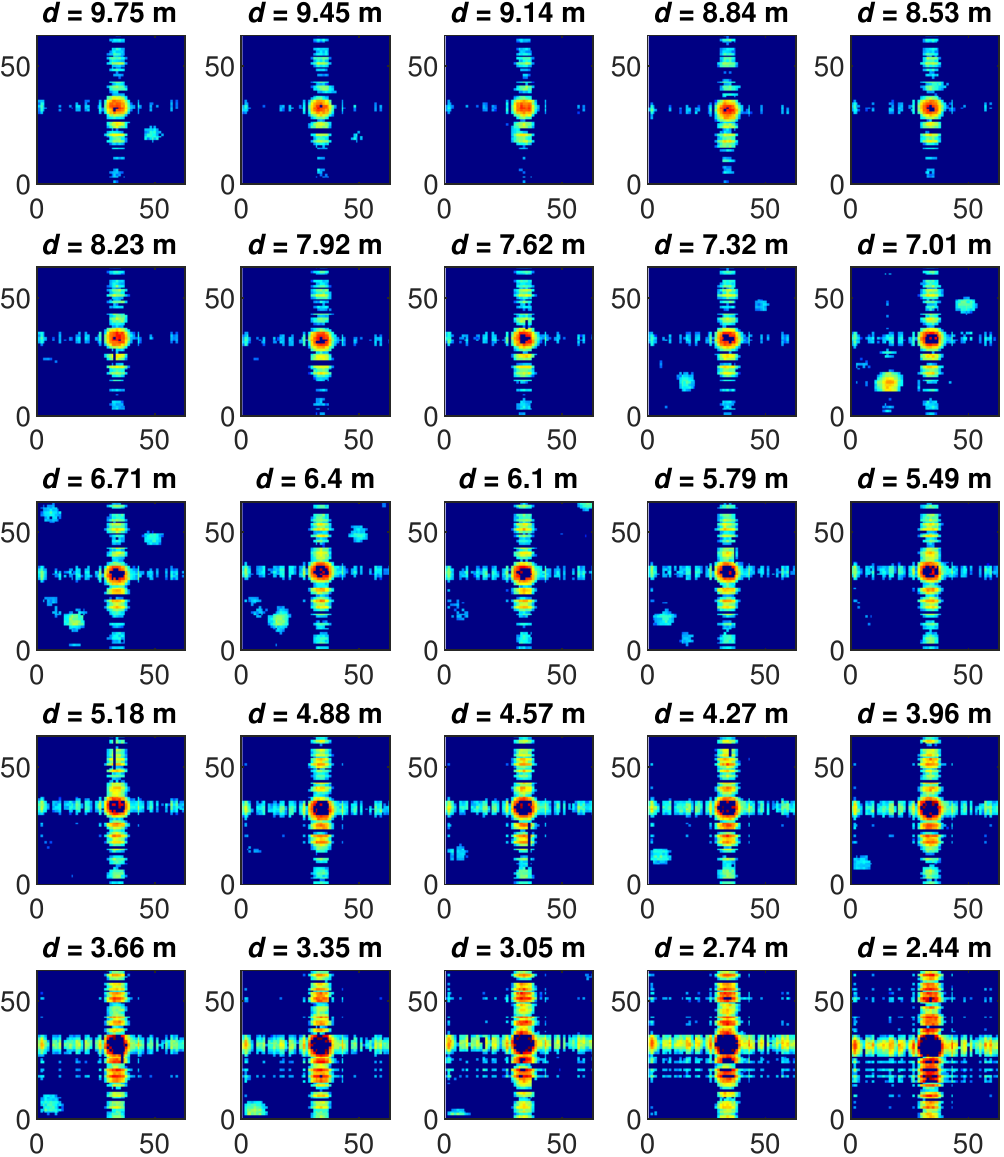}
		}}~~
		\subfloat[$\carrierFrequency= 65.34$~GHz.]{{\includegraphics[width =3.2in]{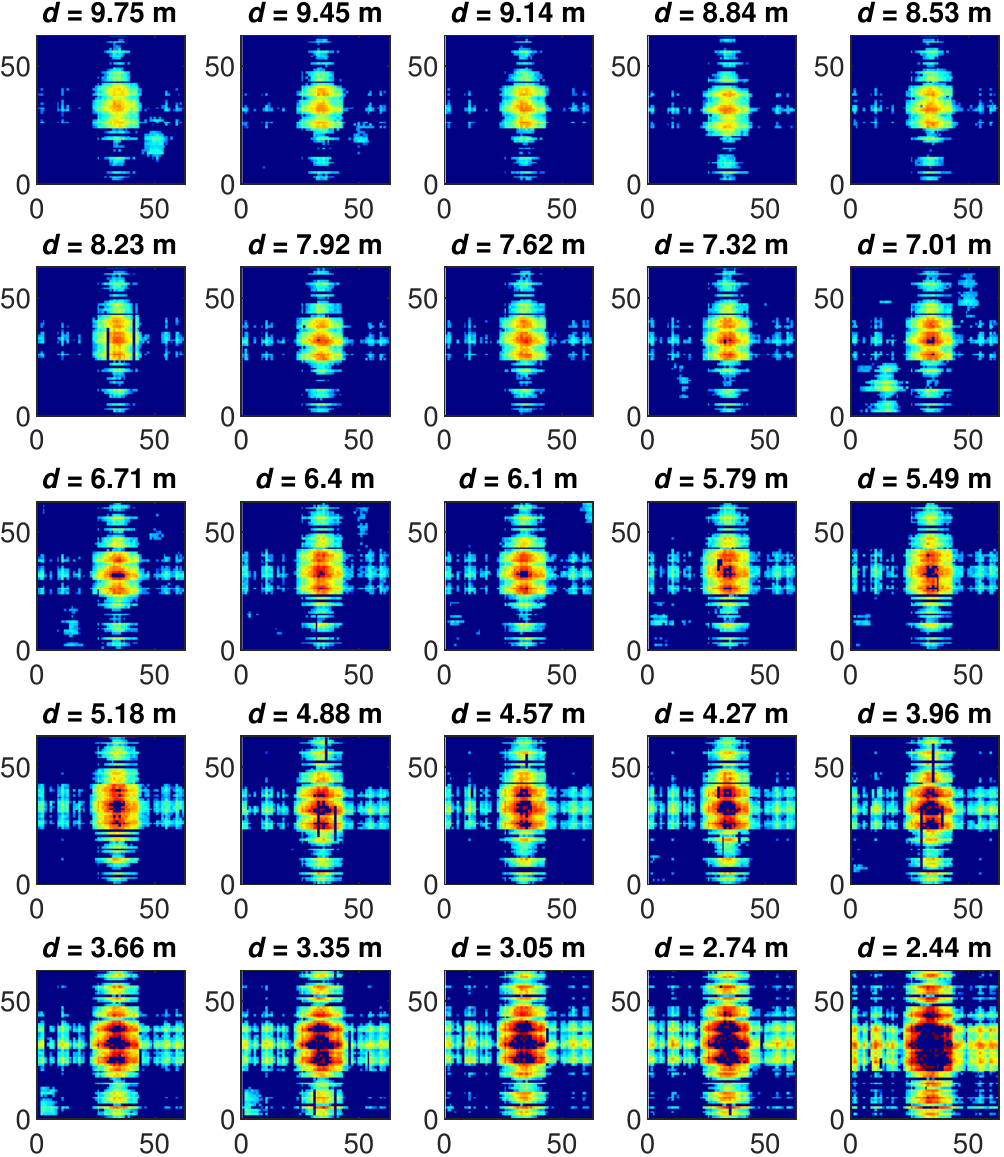}
		}}
		\caption{SNR for a given RX AWV index (x-axis) and TX AWV index (y-axis) for different link distances.  Maximum  and minimum SNRs are 30 dB (red) and 0 dB (blue), respectively.  For both carrier frequencies, we use the AWVs designed for 60.48 GHz. Hence, due to the mismatch between the carrier frequency used for the design of the AWVs and the carrier frequency, the beam pattern is not focused at 65.34 GHz and the antenna sidelobes allow communications.}
		\label{fig:snrmatrix}
		\vspace{-2mm}
	\end{figure*}
	In \figurename~\ref{fig:dataPoint}, as an example, we show the received IQ data samples and the measured \ac{CFR} and \ac{CIR} for the 29th transfer (60.48 GHz) and 20th transfer (65.34 GHz)  when the RX AWV is 31 and the link distance is 9.75~m. After decoding the \ac{PPDU}, TX AWV index is detected as 50 for both cases and the measured SNRs are calculated as 25.76 dB and 24.03 dB for $\carrierFrequency= 60.48$~GHz and $\carrierFrequency= 65.34$~GHz, respectively. The measured \acp{CFR} are relatively flat and similar, where the gap in the measurement is due to the null DC subcarriers. At the beginning of the IQ data, we also observe the test waveform.
	
	In \figurename~\ref{fig:snrmatrix}, we provide the SNRs for a given TX-RX AWV index pair at different locations. We can infer the following. First, the SNR can reach up to 30 dB when the beams are well-aligned, e.g., when the RX AWV and TX AWV indices are around 32, i.e., $0$ degrees. Second, if the received signal is powerful, the receiver may not be able to decode the PPDU due to saturation. We use fixed TX and RX gains in the experiment. Third, the link can still be maintained over a reflection. For instance, for $\carrierFrequency=60.48$~GHz and $\linkDistance=7.01$~m, it is possible to maintain the link if both TX AWV and RX AWV indices are set to 13, likely over a reflection due to the metal cabinets. Fourth, we observe a large difference in the SNR matrices when $\carrierFrequency$ is switched to $65.34$~GHz. Since we still use the \acp{AWV} for $\carrierFrequency=60.48$~GHz, the beams are not focused for $\carrierFrequency=65.34$~GHz. While the mismatch allows the link via the antenna side lobes, it becomes more blind to the reflections as the power is dispersed to the different angles.

	\section{Concluding Remarks}
	In this study, we propose a mmWave SDR solution for  experimentation in the 60~GHz band and introduce \ac{WTR} and a buffering approach for discontinuous transmission to achieve a flexible \ac{CC}-based baseband signal processing. We also generate a new dataset based on a beam sweeping experiment. The main advantages of the proposed SDR are that it is low-cost, portable, and easy to construct. Also, we provide the source code publicly for further development. In future work, we will extend the proposed SDR by integrating and testing it with \acp{UAV} in the NSF AERPAW platform at 28~GHz. In particular, we will focus on beam steering experiments to support links with a mobile UAV. Development of burst type of transmission along with WTR, the potential use of onboard RAM for longer recording, and leveraging the UAV mobility for achieving wide-angle beam sweeping are several possible areas that we will investigate. 
	We will develop a digital twin, emulating the UAVs in conjunction with the proposed SDRs and develop suitable channel models for this environment.
	
	
	\section*{Acknowledgment}
	The authors would like to thank Brian A. Floyd for his contributions to the discussions on the RF design.
	
	\acresetall

	\bibliographystyle{IEEEtran}
	\bibliography{references}
	
\end{document}

%% file: variables.tex
\def\realNumbers{\mathbb{R}}
\def\figuresize{3.5in}

\def\expectationOperator[#1][#2]{{\mathbb{E}_{#2}}\left[#1\right]}
\def\indicatorFunction[#1]{\mathbb{I}\left[{#1}\right]}
\def\probability[#1]{\mathbb{P}\left({#1}\right)}
\def\complexGaussian[#1][#2]{\mathcal{CN}({#1,#2})}
\def\gaussian[#1][#2]{\mathcal{N}({#1,#2})}

\def\normalPDF[#1]{\phi\left(#1\right)}
\def\normalCDF[#1]{\Phi\left(#1\right)}

\def\dataSymbols[#1]{d_{#1}}

\def\linkDistance{d}

\def\carrierFrequency{f_{\rm c}}

\def\oneVector[#1]{\textbf{\textrm{1}}_{#1}}
\def\zeroVector[#1]{\textbf{\textrm{0}}_{#1}}
\def\identityMatrix[#1]{{\textbf{\textrm{I}}_{#1}}}

\def\symbolVector[#1]{\textbf{\textrm{d}}}
\def\transmittedVector[#1]{\textbf{\textrm{t}}}

\def\idftMatrix[#1]{\textbf{\textrm{F}}_{#1}^{\rm H}}
\def\dftMatrix[#1]{\textbf{\textrm{F}}_{#1}}
\def\transformPrecoder[#1]{\textbf{\textrm{T}}_{#1}}
\def\transformDecoder[#1]{\textbf{\textrm{T}}_{#1}^{\rm H}}
\def\dftPrecoder[#1]{\textbf{\textrm{D}}_{#1}}
\def\dftDecoder[#1]{\textbf{\textrm{D}}_{#1}^{\rm H}}

\def\fref{f_{\rm ref}}
\def\fpl{f_{\rm PL}}
\def\sampleRate{f_{\rm sample}}

\def\indexSample{n}
\def\ssSync{\textbf{x}_{\rm SYNC}}

\def\signalTXUL[#1]{\textbf{x}_{{\rm UL},#1}}

\def\iqTXinput[#1]{x_{\rm tx}[#1]}
\def\iqRXinput[#1]{x_{#1}}

\def\rollOff{\beta}
\def\Nup{N_{\rm up}}
\def\rrc[#1]{h_{\rm RRC}[#1]}
\def\golaySequnce{\textbf{g}}
\def\golaySequnceEle[#1]{g_{#1}}
\def\xcorr[#1]{\rho_{#1}}
\def\xcorrParallel[#1][#2]{\rho_{#1,#2}}
\def\acorr[#1]{r_{#1}}
\def\metric[#1]{m_{#1}}
\def\signalOne[#1]{\textbf{x}_{#1}}
\def\signalTwo{\textbf{b}}
\def\signalTwoEle[#1]{b_{#1}}

\def\indexCorr{k}
\def\indexParallel{l}

\def\FPGAcontroller{{\tt FPGActrl}}
\def\EVKcontroller{{\tt EVKctrl}}
\def\API{{\tt API}}

\def\monitor{{\tt monitorIP}}
\def\detector{{\tt detectorIP}}
\def\rxPacketGeneretor{{\tt rxPctGenIP}}
\def\txPacketGeneretor{{\tt txPctGenIP}}

\def\monitor{{\rm IP}_{\rm monitor}}
\def\detector{{\rm IP}_{\rm detector}}
\def\rxPacketGeneretor{{\rm IP}_{\rm packetGenRX}}
\def\txPacketGeneretor{{\rm IP}_{\rm packetGenTX}}

\def\depthFIFO{D_{\rm FIFO}}
\def\widthFIFO{W_{\rm FIFO}}
\def\conversionRate{R_{\rm}}

\def\transferSizeTX{L_{\rm tx}}
\def\triggerSoftTX{t_{\rm tx}}
\def\sizeIQdataTX{S_{\rm tx}}

\def\modeRX{m_{\rm rx}}
\def\transferSizeRX{L_{\rm rx}}
\def\triggerSoftRX{t_{\rm rx, s}}
\def\triggerWTR{t_{\rm rx, w}}
\def\enableTransferRX{e_{\rm rx}}
\def\sizeIQdataRX{S_{\rm rx}}

\def\cntFIFOADC{D_{\rm adc,I}}
\def\thresholdForStopRX{D_{\rm th}}
\def\cntTransferRX{N_{\rm trans}}
\def\cntDetect{N_{\rm detect}}
\def\transferCntResetRX{r_{\rm trans}}

%% file: acronyms.tex
\acrodef{WSN}{wireless sensor network}
\acrodef{SN}{sensor node}
\acrodef{FC}{fusion center}
\acrodef{MAC}{multiple-access channel}
\acrodef{FL}{federated learning}
\acrodef{ED}{edge device}
\acrodef{CS}{compressed sensing}
\acrodef{ES}{edge server}
\acrodef{AXI}{advanced extensible interface}
\acrodef{CC}{companion computer}
\acrodef{CFO}{carrier frequency offset}
\acrodef{RRC}{root-raised cosine}
\acrodef{CHEST}{channel estimation}
\acrodef{PPDU}{physical layer protocol data unit}
\acrodef{DC}{direct current}
\acrodef{CRC}{cyclic redundancy check}
\acrodef{USRP}{universal software-radio peripheral}
\acrodef{ZC}{Zadoff-Chu}
\acrodef{TCP/IP}{transmission control protocol/internet protocol}
\acrodef{FIR}{finite impulse response}
\acrodef{WLAN}{wireless local-area network}

\acrodef{AERPAW}{Aerial Experimentation and Research Platform for Advanced Wireless}
\acrodef{AP}{access point}

\acrodef{STLC}{space-time line code}
\acrodef{CCI}{co-channel interference}
\acrodef{CSIT}[CSIT]{\ac{CSI} at the transmitter}
\acrodef{CSIR}[CSIR]{\ac{CSI} at the receiver}
\acrodef{MIMO}{multiple-input-multiple-output}
\acrodef{PC}{phase correction}
\acrodef{ZF}{zero-forcing}
\acrodef{ANOVA}{analysis of variance}
\acrodef{SDR}{software-defined radio}
\acrodef{PS}{processing system}
\acrodef{SS}{soft synchronization}
\acrodef{IQ}{in-phase/quadrature}

\acrodef{IP}{intellectual property}

\acrodef{DMA}{direct-memory access}
\acrodef{RAM}{random access memory}
\acrodef{UE}{user equipment}
\acrodef{STA}{station}
\acrodef{BS}{base station}
\acrodef{MS}{mobile station}
\acrodef{MSE}{mean squared error}
\acrodef{TDMA}{time-domain multiple access}
\acrodef{GPS}{Global Positioning System}

\acrodef{FPGA}{field-programmable gate array}

\acrodef{FSK}{frequency-shift keying}
\acrodef{PPM}{pulse-position modulation}
\acrodef{PAM}{pulse-amplitude modulation}

\acrodef{MRC}{maximum-ratio combining}
\acrodef{HP}{hard-coded participation}
\acrodef{HPA}{hard-coded participation with absentees}
\acrodef{SP}{soft-coded participation}
\acrodef{FSK-MV}{\ac{FSK}-based \ac{MV}}
\acrodef{RF}{radio-frequency}
\acrodef{MF}{matched filter}
\acrodef{PPM}{pulse-position modulation}
\acrodef{CSK}{chirp-shift keying}
\acrodef{PPM-MV}[PPM-MV]{\ac{PPM}-based \ac{MV}}
\acrodef{DFT-s-OFDM}{\ac{DFT}-spread \ac{OFDM}}
\acrodef{SC}{single-carrier}
\acrodef{SGD}{stochastic gradient descent}
\acrodef{signSGD}{sign stochastic gradient descent}

\acrodef{SL}{split learning}
\acrodef{SNR}{signal-to-noise ratio}
\acrodef{RMSE}{root-mean-square error}
\acrodef{OFDM}{orthogonal frequency division multiplexing}
\acrodef{DFT}{discrete Fourier transform}
\acrodef{PSK}{phase-shift keying}
\acrodef{QAM}{quadrature amplitude modulation}
\acrodef{QPSK}{quadrature phase-shift keying}
\acrodef{PMEPR}{peak-to-mean envelope power ratio}
\acrodef{BER}{bit-error ratio}
\acrodef{SNR}{signal-to-noise ratio}
\acrodef{PSD}{power spectral density}
\acrodef{SE}{spectral efficiency}
\acrodef{CP}{cyclic prefix}
\acrodef{AWGN}{additive white Gaussian noise}
\acrodef{CFR}{channel frequency response}
\acrodef{CIR}{channel impulse response}
\acrodef{MMSE}{minimum mean square error}
\acrodef{LMMSE}{linear minimum mean square error}
\acrodef{BPSK}{binary phase shift keying}
\acrodef{BLER}{block-error rate}
\acrodef{ML}{maximum likelihood}
\acrodef{PHY}{physical layer}
\acrodef{PA}{power amplifier}
\acrodef{IDFT}{inverse DFT}
\acrodef{DoF}{degrees-of-freedom}
\acrodef{IoT}{Internet-of-Things}
\acrodef{FDE}{frequency-domain equalization}
\acrodef{RF}{radio-frequency}
\acrodef{IM}{index modulation}
\acrodef{BS}{base station}
\acrodef{MF}{matched filter}
\acrodef{PPM}{pulse-position modulation}

\acrodef{MSE}{mean-square error}
\acrodef{MRT}{maximum-ratio transmission}
\acrodef{ERC}{equal-ratio combining}
\acrodef{BAA}{broadband analog aggregation}
\acrodef{OBDA}{one-bit broadband digital aggregation}
\acrodef{FEEL}{federated edge learning}
\acrodef{FL}{federated learning}
\acrodef{ED}{edge device}
\acrodef{ES}{edge server}
\acrodef{UL}{uplink}
\acrodef{DL}{downlink}
\acrodef{OAC}{over-the-air computation}
\acrodef{TCI}{truncated-channel inversion}
\acrodef{MV}{majority vote}
\acrodef{CNN}{convolution neural network}
\acrodef{ReLU}{rectified-linear unit}
\acrodef{CSI}{channel state information}
\acrodef{PAPR}{peak-to-average power ratio}
\acrodef{SC}{single-carrier}
\acrodef{iid}[IID]{independent and identically distributed}
\acrodef{RMS}{root-mean-square}
\acrodef{4G}{Fourth Generation}
\acrodef{5G}{Fifth Generation}
\acrodef{NR}{New Radio}
\acrodef{LTE}{Long-Term Evolution}
\acrodef{DFT-s-OFDM}{\ac{DFT}-spread \ac{OFDM}}
\acrodef{OFDMA}{orthogonal frequency division multiple access}
\acrodef{HARQ}{hybrid automatic repeat request}
\acrodef{D2D}{Device-to-Device}
\acrodef{NOMA}{non-orthogonal multiple access}

\acrodef{WTR}{waveform-triggered reception}
\acrodef{STR}{software-triggered reception}
\acrodef{DAC}{digital-to-analog converter}
\acrodef{ADC}{analog-to-digital converter}
\acrodef{mmWave}{millimeter-wave}
\acrodef{UAV}{unmanned air vehicle}

\def\mmWaveSDR{mmWave-SDR}

\acrodef{mmWaveSDR}[\mmWaveSDR]{\acl{mmWave} \acl{SDR}}

\acrodef{MTS}{multi-tile synchronization}
\acrodef{PAA}{phased-antenna array}
\acrodef{PL}{programmable logic}
\acrodef{PS}{programmable system}
\acrodef{SPI}{serial peripheral interface}
\acrodef{USB}{universal serial bus}
\acrodef{AGC}{automatic gain control}
\acrodef{FIFO}{first-in-first-out}
\acrodef{API}{application programming interface}
\acrodef{LO}{local oscillator}
\acrodef{AWV}{antenna weighting vector}
\acrodef{XRFDC}{Xilinx RF data converter}
\acrodef{PPD}{poly-phase detector}